\documentclass[12pt]{iopart}
\usepackage[dvips]{graphicx}

\newcommand \beq{\begin{eqnarray}}
\newcommand \eeq{\end{eqnarray}}

\def\simge{\mathrel{%
       \rlap{\raise 0.511ex \hbox{$>$}}{\lower 0.511ex \hbox{$\sim$}}}}
\def\simle{\mathrel{
       \rlap{\raise 0.511ex \hbox{$<$}}{\lower 0.511ex \hbox{$\sim$}}}}

\begin{document}

\title{The axial anomaly and the phases of dense QCD}

\author{Gordon Baym,$^1$ Tetsuo Hatsuda,$^2$ Motoi Tachibana,$^{3}$ and
Naoki Yamamoto$^{2}$ }

\address{$^{1}$Department of Physics, University of Illinois, 1110 W. Green St.,
Urbana, Illinois 61801\\
$^{2}$Department of Physics, The University of Tokyo, Tokyo 113-0033,
Japan\\
$^{3}$Department of Physics, Saga University, Saga 840-8502, Japan\\}

\ead{gbaym@uiuc.edu}

\begin{abstract}

    The QCD axial anomaly, by coupling the chiral condensate and BCS pairing
fields of quarks in dense matter, leads to a new critical point in the QCD
phase diagram \cite{HTYB,chiral2}, which at sufficiently low temperature
should terminate the line of phase transitions between chirally broken
hadronic matter and color superconducting quark matter.  The critical point
indicates that matter at low temperature should cross over smoothly from the
hadronic to the quark phase, as suggested earlier on the basis of symmetry.
We review here the arguments, based on a general Ginzburg-Landau effective
Lagrangian, for the existence of the new critical point, as well as discuss
possible connections between the QCD phase structure and the BEC-BCS crossover
in ultracold trapped atomic fermion systems at unitarity. and implications for
the presence of quark matter in neutron stars.

\end{abstract}


\maketitle

\section{Introduction}

    Understanding the properties of matter under extreme conditions has been
an important motivation for carrying out ultrarelativistic heavy ion
collisions \cite{gbrhic}.  The general features of dense matter can be
summarized in the QCD phase diagram, various aspects of which have been
discussed by several speakers here, e.g., [4-7].  Crucial refinements of the
phase diagram have been the high temperature critical point -- point C in
Fig.~1 -- (reviewed in \cite{MPL}), and the phases of color superconducting
quark matter at low temperature and high density \cite{alford}.

\begin{figure}[h]
\begin{center}
\includegraphics[width=11.0cm]{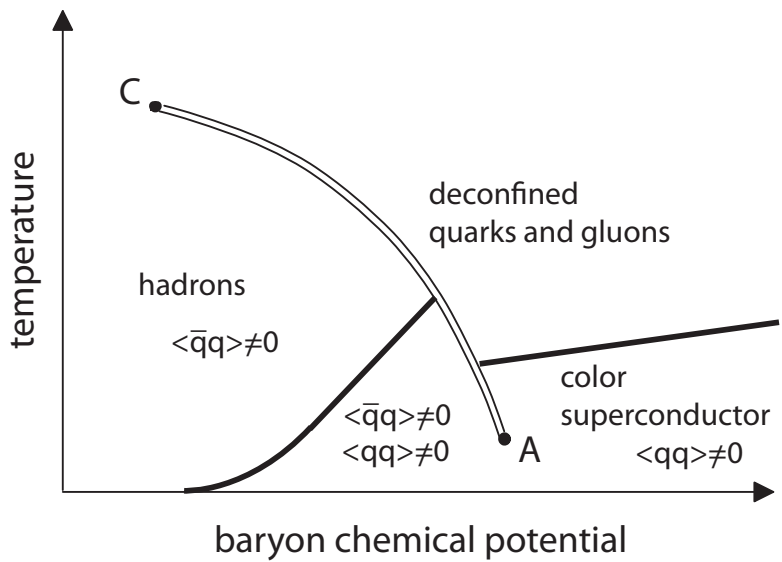}
\end{center}
\begin{center}
\caption{Schematic phase diagram of dense QCD with two light up and down
quarks and a medium heavy strange quark, showing the new critical point, A, at
low temperature, and regions of broken chiral symmetry, $\langle \bar q
q\rangle \ne 0$, and regions of diquark pairing, $\langle q q\rangle \ne 0$.
In the region to the left of A broken chiral symmetry and quark pairing
coexist.}
\end{center}
\label{fig1}
\end{figure}

    Here we discuss how the U(1)$_A$ axial anomaly, by coupling two competing
phenomena -- quark-anti-quark pairing, characterized by a chiral condensate
$\Phi \sim \langle \bar{q}q \rangle$, and quark-quark pairing, characterized
by a diquark condensate $d \sim \langle qq \rangle$ -- induces a second
critical point in the phase diagram, point A in Fig.~1.  The anomaly leads to
an effective six quark 't Hooft interaction, $\sim {\rm det}_{ij}\langle \bar
q^j_R q^i_L\rangle$, where $i$ and $j$ are flavor indices, and $R$ and $L$
denote quark chiralities; this interaction in turn implies couplings $\sim
\Phi^3$ and $\Phi d^\dagger d$; the latter coupling, which favors coexistence
of chiral symmetry breaking and BCS quark pairing, is responsible for the new
critical point.

\section{Ginzburg-Landau effective Lagrangian}

    Since in the neighborhood of chiral symmetry breaking and quark pairing
transitions $d$ and $\Phi$ are small, we construct the free energy by means of
a model-independent Ginzburg-Landau effective Lagrangian developed in powers
of $d$ and $\Phi$.  Owing to space limitations, we make discuss only the
simple case that for three massless flavors, $\Phi = {\rm
diag}(\sigma,\sigma,\sigma)$, and the pairing field is in the color-flavor
locked phase \cite{alford}, $d_L=-d_R={\rm diag}(d,d,d)$, with real order
parameters $\sigma$ and $d$.  The minus sign corresponds to the vacuum having
even parity.  See Refs.~\cite{HTYB} and \cite{chiral2} for full details.  The
free energy, measured with respect to that for $\sigma=d=0$, is a sum,
$\Omega_{3F} = \Omega_0+\Omega_\chi + \Omega_d + \Omega_{\chi d}$, of
contributions from the chiral field, the pairing field, and their interaction,
respectively.  The chiral term has the form \cite{PW84},
\beq
 \Omega_\chi = \frac{a}{2} \sigma^2
 - \frac{c}{3} \sigma^3 + \frac{b}{4}\sigma^4 + \frac{f}{6}\sigma^6;
 \label{chi}
\eeq
the $c$ term arises from the axial anomaly.  The system breaks chiral
symmetry when $a<2c^2/9b$, for $b>0$.  The $f\sigma^6/6$ is needed to
stabilize the system should $b$ be negative.  The pairing free energy
\cite{IB} is
\beq
\label{d}
\Omega_{d} = \frac{\alpha}{2} d^2 + \frac{\beta}{4} d^4;
\eeq
the system pairs when $\alpha< 0$.  Finally, the interaction between the
fields, to fourth order in the fields, is
\beq
\label{int}
\Omega_{\chi d} = - {\gamma} d^2 \sigma + {\lambda} d^2 \sigma^2.
\eeq
The $\gamma$ term, which breaks the $U(1)_A$ symmetry down to $Z(6)$,
originates from the axial anomaly.\footnote {The general Ginzburg-Landau
Lagrangian \cite{HTYB,chiral2}, aside from the two terms arising from the
anomaly, is constructed to be fully invariant under the group ${\cal G} = {\rm
SU(3)_L \times SU(3)_R \times U(1)_B \times U(1)_A \times SU(3)_C}$.} The 't
Hooft interaction implies that the coefficient $c$ of the $\sigma^3$ term, and
$\gamma$ have the same (positive) sign and similar magnitude.  The term
$\gamma$ term encourages the coexistence of the pairing and chiral fields, and
is eventually responsible for the new critical point; a non-zero $d^2$ acts as
external field for $\sigma$, washing out the first order transition for large
$d^2$, as in a magnetic system in external field.  On the other hand, in the
presence of the $\gamma$ term, the $\lambda d^2 \sigma^2$ plays little
quantitative role, since $\lambda/\beta$ is expected to be $\ll 1$
\cite{HTYB}.

    The phase structure for three massless flavors in the $a,\alpha$ plane
implied by $\Omega_{3F} = \Omega_0+\Omega_\chi + \Omega_d + \Omega_{\chi d}$,
for $b>0$, is shown in the left panel of Fig.~2.  The ``normal"
(Nambu-Goldstone like) region to the upper left has broken chiral symmetry,
while in the ``QGP" region in the upper right, chiral symmetry is restored; in
neither of these regions are the quarks paired.  The point D is always at
positive $\alpha$, since a non-zero $\sigma$ enhances the pairing gap, via the
$- {\gamma} d^2 \sigma$ in Eq.~(\ref{int}), and $\sigma$ is larger at smaller
chemical potential \cite{HTYB,chiral2}.  The region to the lower right is
color superconducting.  As a consequence of the $\gamma$ coupling, the double
line (which for $\gamma = 0$ runs parallel to the $\alpha$ axis down to
$-\infty$) terminates at a critical point -- always at positive $a>0$ and
negative $\alpha$ -- connecting the color superconducting region in the lower
right continuously to a coexistence region to the left.  The coexistence
region is characterized by both chiral symmetry breaking and pairing.  As a
consequence of the critical point, the color superconducting region also has a
non-zero chiral field, $\sigma$.  Across phase transition marked by the double
line, $\sigma$ decreases discontinuously with increasing $a$.

    Similarly for two massless flavors ($m_s=\infty$), we write $\Phi ={\rm
diag}(\sigma,\sigma,0)$ and $d_L=-d_R={\rm diag}(0,0,d)$, corresponding to the
2SC state of color superconductivity.  With this color-flavor structure, the
cubic terms in $\sigma$ and $d$ are identically zero, leading to a phase
diagram with a much simpler structure than for three massless flavors.  The
free energy is
\beq
\label{eq:nf2-model}
 \Omega_{2F} =
  \left( \frac{a}{2} \sigma^2 + \frac{b}{4} \sigma^4
 + \frac{f}{6} \sigma^6 \right)
 +\left(  \frac{\alpha}{2} {d}^2 + \frac{\beta}{4} {d}^4 \right)
 + {\lambda} {d}^2 \sigma^2,
\eeq
where the coefficients are the same as in the three flavor case times
numerical factors of order unity.

    The right panel of Fig.~2 shows the phase structure for two massless
flavors in the $a,\alpha$ plane, for $b >0$.  The solid lines are second order
transitions.  Unlike for three flavors, there is no critical point; rather in
the color superconducting phase, $\sigma = 0$, owing to the absence of the
$\gamma d^2\sigma$ coupling.

\begin{figure}[h]
\begin{center}
\includegraphics[width=15.0cm]{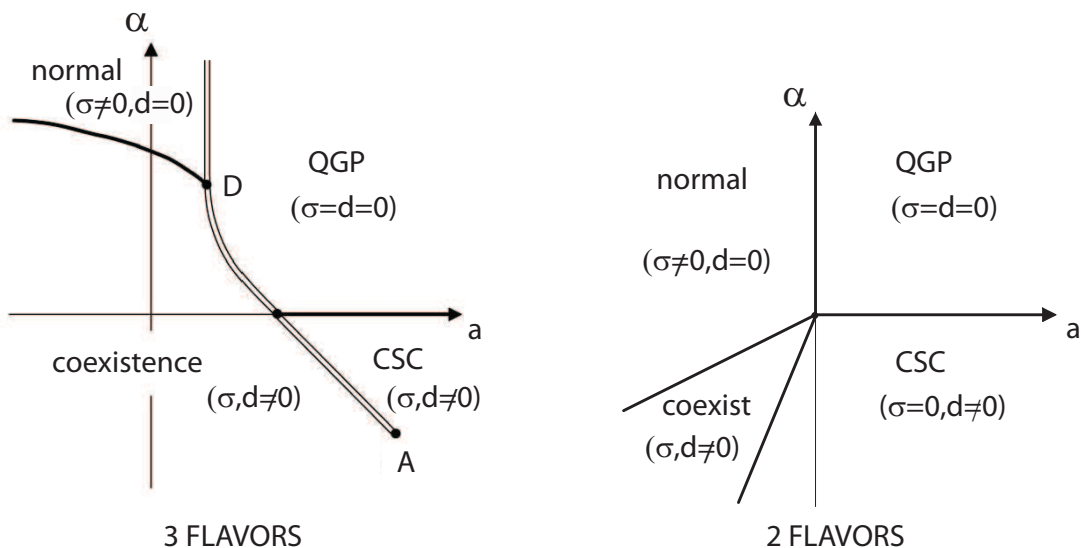}
\end{center}
\begin{center}
\caption{The phase structure in the $a,\alpha$ plane for three massless
flavors (left panel) and two massless flavors (right panel).  The critical
point for three massless flavors is denoted by A.}
\end{center}
\label{fig:32F}
\end{figure}

    In the presence of the axial anomaly, the coexistence and the color
superconducting phases have same symmetry, allowing a crossover between them.
On the other hand, the normal and coexistence phases realize U(1)$_{\rm B}$
differently and the boundary between these regions, the solid line in both
panels of Fig.~2, is sharp.

    Mapping the phase diagram from the $a,\alpha$ plane to the baryon chemical
potential -- temperature plane requires a dynamical picture to calculate, for
realistic quark masses, the Ginzburg-Landau parameters, and to treat the full
interplay between chiral symmetry breaking, quark pairing, and confinement,
e.g., via lattice simulations or effective theories of confinement
\cite{pnjl}.  Generally, we expect $a$ and $\alpha$ to decrease from positive
to negative with decreasing temperature.  We further expect the situation with
realistic quark masses to be intermediate between the two and three flavor
massless cases, and thus expect the schematic phase diagram for realistic
quark masses shown in Fig.~1.  NJL estimates of the Ginzburg-Landau parameters
\cite{chiral2} indicate that the critical values of $a$ and $\alpha$ at the
critical point A are reached at finite temperature, so that the critical point
occurs at non-zero temperature.  With increasing strange quark mass, the
critical point A is expected to move to lower temperature.  The solid line on
the left demarcating the normal and coexistence regions must touch the double
line at higher temperature than the color superconducting region.  The new
critical point is at temperature below the critical temperature for onset of
color pairing, and thus would be difficult to study at RHIC and LHC; however,
it may be accessible at FAIR.

    The three flavor coexistence region to the left in which both $d$ and
$\sigma$ are non-zero is an intermediate state between color-paired quark
matter and hadronic matter.  The details of this state, which depend crucially
on understanding interplay with confinement, remain to be delineated; below we
discuss suggestive similarities of this structure to the BEC-BCS crossover in
ultracold atomic systems.  An essential feature of the schematic phase diagram
here is the continuity between the coexistence region and the color-paired
region.  The Ginzburg-Landau picture points towards an explicit realization of
``hadron-quark continuity" \cite{SW99}.

\section{BEC-BCS crossover and the deconfinement transition}

    Although the energy scales in quark-gluon plasmas are some 21 orders of
magnitude greater than in ultracold trapped atomic systems (on the
sub-microkelvin scale), strongly interacting atomic fermion systems share
certain intriguing overlaps with quark-gluon plasmas, e.g., first viscosities
approaching the KSS bound \cite{visc}, and possible BCS pairing in imbalanced
fermionic gases \cite{imbalance}.  Particular relevant, the crossover from the
coexistence region to quarks at finite baryon density, Fig.~1, and the
crossover from Bose-Einstein condensation of molecules to BCS-paired
superfluids in two component cold fermionic gases, Fig.~3, bear suggestive
similarities, opening possible new approaches to the deconfinement transition.

\begin{figure}[h]
\begin{center}
\vspace{1cm}
\includegraphics[width=12.0cm]{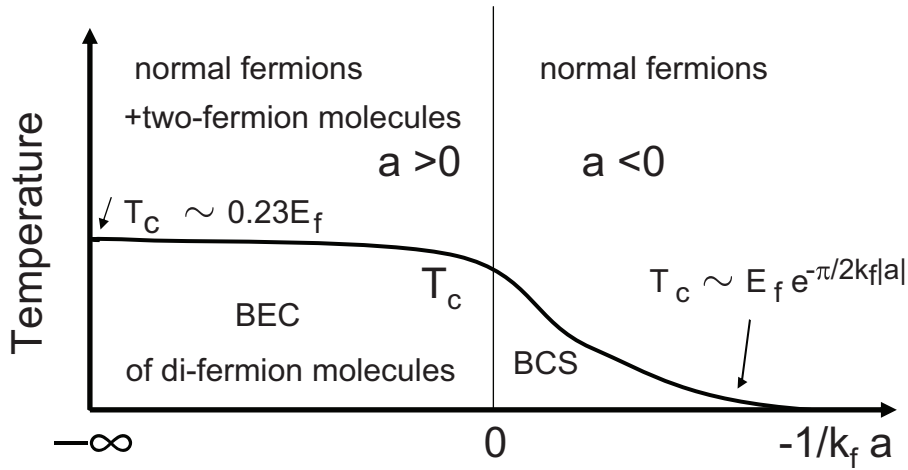}
\vspace{-1.5cm}
\caption{ Phase diagram of ultracold atomic fermions as a
function of the negative of the inverse scattering length $a$, in units of the
Fermi momentum.  The continuous curve is the transition temperature for
condensation, approaching the weakly interacting Bose-Einstein transition
temperature to the left and the BCS transition temperature to the right;
$E_f$ is the free particle Fermi energy.  The transition between the BEC and
BCS regions is a smooth crossover.}
\end{center}
\end{figure}

    Figure 3 shows the finite temperature phase diagram of a gas of two
equally populated hyperfine states of ultracold fermionic atoms.  The atoms
interact with an attractive potential, with an s-wave scattering length, $a$,
which can be tuned in experiment from large positive to large negative values
by means of a Feshbach resonance in an external magnetic field.  The
horizontal axis is $-1/k_fa$, where $k_f$ is the Fermi momentum of the gas.
At small positive $a$, the system has strongly bound states -- di-fermion
molecules -- which undergo Bose condensation at low temperature, while at
small negative $a$ the fermions become BCS paired at low temperatures.
Through the Feshbach resonance, where $|a|$ becomes large compared with the
interparticle spacing, the strongly interacting system is scale free, similar
to a quark-gluon plasma at high temperature.  Remarkably, in the region
between these two extremes the system undergoes a gradual crossover from a BEC
superfluid of molecules to a BCS paired superfluid \cite{crossover}.  Nothing
dramatic happens as one goes through the limit $|a|\to\infty$; rather, the
molecules continuously expand in size from tightly bound on the BEC side to
widely spaced pairs in the BCS regime, while the system remains superfluid
\cite{martinvort}.

    The correspondences with the phase diagram of dense QCD matter, Fig. 1,
are apparent.  At very large baryon chemical potential, $\mu$, and low
temperature the matter is BCS-paired, as are the atomic systems at large
$-1/k_fa$, and low temperature.  As the temperature increases the atomic
system becomes a fluid of normal unpaired fermions, while the QCD system
becomes an unpaired plasma.  Furthermore, with decreasing $-1/k_fa$ the atomic
system continuously transforms into a Bose-Einstein condensate of di-fermion
molecules; similarly QCD matter at low temperature goes from the color-paired
states at large $\mu$ smoothly under the critical point, to the coexistence
state in which both $\langle qq\rangle$ and $\langle \bar q q\rangle$ are
non-zero.

    Were the color symmetry group only SU(2), one could make a strong analogy
between the atomic and quark systems -- diquark bosonic baryons would go over
into BCS pairs at higher densities, just as the atomic molecules expand into
BCS pairs through the Feshbach resonance.  However in SU(3)$_{\rm C}$, the
system must develop strong three-quark correlations at low densities from the
two-quark correlations at high densities.  The atomic systems suggest that
these correlations can enter through (2SC) BCS pairs shrinking into diquarks
as the system enters the coexistence region, forming a Bose-Einstein
condensate of diquarks (the extension of the 2SC phase); and that at the solid
line in Fig. 1, the diquarks bind to the unpaired quarks to form nucleons.
However, construction of such a picture requires understanding more fully the
effect of residual interactions on the momentum distributions of the
quarks \cite{diquark}.

\section{Dense quark matter and neutrons stars}

    The canonical phenomenological approach to deduce the onset of quark
matter in high density nuclear matter is to calculate the equations of state
of hadronic matter and quark matter; when the thermodynamic potential for
quark matter becomes lower than that for hadronic matter, the system is
presumed to undergo a first order phase transition to the quark state, e.g.,
\cite{chin}.  Typical nuclear matter equations of state are based on
nucleon-nucleon interactions deduced from scattering experiment, together
with three body forces constrained by the light nuclei, e.g., \cite{vijay}.
From such an approach one would conclude that quark matter first appears at
baryon densities, $n$, of order 7-10 times nuclear matter density, $n_0$, and
thus the observation of high mass neutron stars (e.g., Vela X-1, Cyg X-2,
where $M$ could be as large as $1.8 M_\odot$) would practically preclude the
existence of quark matter cores in neutron stars.

    However, an equation of state based on nucleon interactions alone, while
accurately describing neutron star matter in the neighborhood of nuclear
matter density, has fundamental limitations.  Beyond a few times $n_0$ one
should not expect that the forces between particles can be described in terms
of static few-body potentials; since the characteristic range of the nuclear
forces is $\sim 1/2m_\pi$, the parameter measuring the relative importance of
three and higher body forces is of order $n/(2m_\pi)^3 \sim 0.4n/{\rm
fm}^{-3}$, so that at densities well above $n_0$ a well defined expansion in
terms of two-, three-, or more, body forces no longer exists.  The nucleonic
equation of state furthermore does not take into account the rich variety of
hadronic degrees of freedom that enter with increasing density.  Nor can one
continue to assume that the system at higher densities can even be described
in terms of well-defined ``asymptotic'' laboratory particles.

    More realistically, one should expect a gradual onset of quark degrees of
freedom in dense matter, degrees of freedom not accounted for by nucleons
interacting via static potentials.  Indeed once nucleons overlap considerably
the matter should percolate, opening the possibility of their quark
constituents propagating throughout the system (although near the onset of
percolation valence quarks may prefer to remain bound in triplets, mimicking
nucleons, leaving the matter a color insulator) \cite{perc,satz}.
Furthermore, the transition from hadronic to quark matter at low temperature,
as suggested by Fig.~1, would involve the matter entering the coexistence
phase, possibly a diquark Bose-Einstein condensate, and then smoothly crossing
over to superfluid quark matter.  The relation of the transition to the
coexistence phase and percolation remains an open question.  A firm assessment
of the role of quarks in neutron stars must await a better understanding of
mechanism of quark deconfinement with increasing baryon density.

    \ack This research was supported in part by the Grants-in-Aid of the
Japanese Ministry of Education, Culture, Sports, Science, and Technology
(No.~18540253), and by U.S.~NSF Grants No.~PHY03-55014 and PHY07-01611.
Author GB thanks the University of Tokyo for its kind hospitality as well as
support through the COE.

\end{document}